\documentclass[letterpaper,  
              ]{jacow}
\makeatletter%
	\ifboolexpr{bool{xetex}}
	 {\renewcommand{\Gin@extensions}{.pdf,%
	                    .png,.jpg,.bmp,.pict,.tif,.psd,.mac,.sga,.tga,.gif,%
	                    .eps,.ps,%
	                    }}{}
\makeatother

\ifboolexpr{bool{xetex} or bool{luatex}} 
 {}                                      
 {\usepackage[utf8]{inputenc}}           

\usepackage[USenglish]{babel}			 

\usepackage[final]{pdfpages}
\usepackage{multirow}
\usepackage{ragged2e}

\usepackage{siunitx}

\begin{document}

\title{Spectral Form Function with Applications in Beam Physics}

\author{ Xiujie Deng\thanks{dengxiujie@mail.tsinghua.edu.cn\\ Presented in the 15th Symposium on Accelerator Physics (SAP2025), Urumqi, Xinjiang, China, Sept 2-5, 2025.}, Institute for Advanced Study,
	 Tsinghua University, Beijing, China 
	}
	
\maketitle

\begin{abstract}
   To describe longitudinal fine structure like microbunching within a particle beam, a classical approach is to define a bunching factor which is the Fourier transform of the particle longitudinal density distribution. Such a 1D definition of bunching factor can be generalized to a 6D spectral form function (SFF) to describe more complicated structure in phase space~\cite{Yampolsky2017}. The complex SFF is another complete description of beam in spectral domain and can offer complementary and valuable insight in beam dynamics study which usually invokes the real particle density distribution. The basic property and Fokker-Planck equation of the SFF is presented, along with its solution in a general coupled linear lattice. The example applications of SFF in electron storage ring physics and laser-induced microbunching are presented.

\end{abstract}

\section{Introduction}

Microbunching enables laser-like radiation generation from charged particle beam, and is one of the main driving forces advancing accelerator light sources in the past decades. The most prominent example is free-electron laser.   Usually we mainly care about the longitudinal coordinate of the particles in quantifying the degree of microbunching, since the radiation of a relativistic beam is dominantly in the forward direction. But strictly speaking the 6D particle phase space coordinates can all have an impact on coherent radiation~\cite{DengSpringer2024}. One can even create novel 6D structures in phase space for various purposes, for example to tailor the radiation properties or to control collective beam dynamics. One example is the creation of helical microbunching for light generation with orbit angular momentum~\cite{Hemsing2009Helical}. The classical 1D definition of bunching factor is clearly not sufficient in many applications. This justifies our motivation to investigate such a generalized definition of spectral form function (SFF). But we recognize the potential applications of SFF can be much broader than this original motivation.




\section{Spectral form function}
6D particle state vector: 
$
	{\bf X}\equiv\left(
	\begin{matrix}
		x&
		x'& 
		y&
		y'&
		z&
		\delta
	\end{matrix}
	\right)^{T}.
$
6D spectral vector: ${\bf K}\equiv\left(
\begin{matrix}
	k_{x}&
	k_{x'}&
	k_{y}&
	k_{y'}&
	k_{z}&
	k_{\delta}
\end{matrix}
\right).$
Normalized charge density function: $\psi({\bf X})$ satisfying $\int\psi({\bf X})d{\bf X}=1$, $\psi({\bf X})\geq0$.
Then SFF is defined as:
\begin{equation}
	\begin{aligned}
		&\mathcal{F}({\bf K})\equiv\int\psi({\bf X})e^{-i{\bf K}{\bf X}}d{\bf X}.
	\end{aligned}
\end{equation}
$\psi({\bf X})$ and $\mathcal{F}({\bf K})$ forms a Fourier transform pair
\begin{equation}
	\psi({\bf X})=\frac{1}{(2\pi)^{6}}\int \mathcal{F}({\bf K})e^{i{\bf K}{\bf X}}d{\bf K}.
\end{equation}

The linear symplectic dynamics in an accelerator is dictated by a quadratic Hamiltonian
$
	\mathcal{H}=\frac{{\bf X}^{T}{\bf H}{\bf X}}{2},
$
where ${\bf H}={\bf H}^{T}$. The Hamiltonian equation in matrix form is then
$
	\frac{d{\bf X}}{ds}={\bf S}{\bf H}{\bf X},
$
with ${\bf S}$ the symplectic form. The evolution of particle state vector from the initial point $s_{i}$ to the final point $s_{f}$ can be described by a symplectic transfer matrix according to
$
	{\bf X}(s_{f})={\bf R}(s_{f},s_{i}){\bf X}(s_{i}),
$
with
$
	{\bf R}(s_{f},s_{i})=e^{\int_{s_{i}}^{s_{f}}{\bf S}{\bf H}ds}
$  if ${\bf H}$ is $s$-independent from $s_{i}$ to $s_{f}$.
Correspondingly the transfer matrix for ${\bf K}$ is 
$
	{\bf K}(s_{f})={\bf K}(s_{i}){\bf R}^{-1}(s_{f},s_{i}).
$
From continuity equation
$
\frac{\partial \psi}{\partial s}+\nabla_{\bf X}\cdot\left(\psi\frac{d{\bf X}}{ds}\right)=0
$
where $\nabla_{\bf X}\equiv\left(\frac{\partial}{\partial{\bf X}_{1}},\cdots,\frac{\partial}{\partial{\bf X}_{6}}\right)$
and Hamiltonian equation follows the Liouville equation
\begin{equation}
\frac{d\psi}{ds}=\frac{\partial\psi}{\partial s}+\left[\psi,\mathcal{H}\right]=0,  \frac{d\mathcal{F}}{ds}=\frac{\partial\mathcal{F}}{\partial s}+\left[\mathcal{H}_{\bf K},\mathcal{F}\right]=0,\
\end{equation} 
with $\mathcal{H}_{\bf K}=-\frac{{\bf K}{\bf S}{\bf {H}}{\bf S}{\bf K}^{T}}{2}$, from which follows  
$
\psi({\bf X},s_{f})=\psi({\bf R}^{-1}(s_{f},s_{i}){\bf X},s_{i}),\
\mathcal{F}({\bf K},s_{f})=\mathcal{F}({\bf K}{\bf R}(s_{f},s_{i}),s_{i}).
$ We recognize the work presented in this section has been obtained before by Yampolsky\cite{Yampolsky2017}.

\section{Fokker-Planck Equation}
Now let us add non-symplectic process, like damping and diffusion. Here we simplify the discussion by assuming that the damping coefficients are independent of the particle state vector, for example that of the radiation damping. The equation of motion is now  
\begin{equation}
\frac{d{\bf X}}{ds}=\left({\bf S}{\bf H}{\bf X}+{\bf B}{\bf X}\right)+{\bf \xi}(s),
\end{equation}
with the stochastic process $\xi$ satisfying
$
\int p(\xi) d\xi=1,\
\int \xi_{i} p(\xi_{i}) d\xi_{i}=0,\
\int \xi_{i}(s)\xi_{j} (s') p(\xi_{i}(s),\xi_{j}(s')) d\xi_{i}d\xi_{j}=D_{ij}\delta(s-s'),
$
where $p(\xi)$ is the probability distribution function of $\xi$. We have assumed that the noise is a Gaussian white noise. In the above equation, ${\bf B}$ is responsible for the deterministic damping or antidamping, and $\xi$ for diffusion. Note that here we actually assume that the diffusion is a continuous-diffusion process, instead of a jump-diffusion process whose rigorous description requires the Kramers-Moyal expansion~\cite{Tabar2019Analysis}. The quantum excitation for example is more accurately modeled by a jump-diffusion process. 
Denote ${\bf C}\equiv {\bf S}{\bf H}+{\bf B}$, and note that $\text{Tr}({\bf S}{\bf H})=0$, we can then derive the Fokker-Planck equation for $\psi({\bf X})$
\begin{equation}\label{eq:FPP}
\begin{aligned}
\frac{\partial\psi}{\partial s}+\nabla_{\bf X}\left(\psi{\bf C}{\bf X}\right)=\frac{1}{2}\sum_{i=1}^{6}\sum_{j=1}^{6}D_{ij}\frac{\partial^{2}\psi}{\partial{\bf X}_{i}\partial{\bf X}_{j}}.
\end{aligned}
\end{equation}
The corresponding equation in spectral domain is
\begin{equation}\label{eq:FPF}
\begin{aligned}
\frac{\partial{\mathcal{F}}}{\partial{s}}-({\bf KC})(\nabla_{\bf K}\mathcal{F})&=-\frac{{\bf K}{\bf D}{\bf K}^{T}}{2}\mathcal{F},
\end{aligned}
\end{equation}
where $\nabla_{\bf K}\equiv\left(\frac{\partial}{\partial{{\bf K}_{1}}},\cdots,\frac{\partial}{\partial{{\bf K}_{6}}}\right)^{T}$.
From the right hand side of Eq.~(\ref{eq:FPF}), it is clear that diffusion has a stronger impact on high-frequency bunching, i.e., finer structures in phase space. Also note that if there is no diffusion in phase space, i.e., if ${\bf D}={\bf 0}$, we have $\frac{d\psi}{ds}=\frac{\partial\psi}{\partial s}+(\nabla_{\bf X}\psi)\left({\bf C}{\bf X}\right)=-\text{Tr}({\bf C}){\psi}=-\text{Tr}({\bf B}){\psi}$, and $\frac{d\mathcal{F}}{ds}=\frac{\partial{\mathcal{F}}}{\partial{s}}-({\bf KC})(\nabla_{\bf K}\mathcal{F})=0$. So the information of fine structure in phase space in some sense can only be destroyed by diffusion or stochastic process. The deterministic linear transport with damping or anti-damping can only rotate, shrink or expand the structure.




Now we solve the Fokker-Planck equation. First we assume the accelerator lattice is a piece-wise one and we solve the equation in each piece, in which ${\bf C}$ is a constant matrix. First we diagonalize the matrix ${\bf C}$, 
$
{\bf C}={\bf P}^{-1}{\bf Q}{\bf P},
$
with ${\bf Q}$ a diagonal matrix, and ${\bf Q}_{ii}=\lambda_{i}$ the eigenvalues of ${\bf C}$. Denote ${\bf K}'\equiv{\bf K}{\bf P}^{-1}$ and ${\bf D}'\equiv{\bf P}{\bf D}{\bf P}^{T}$, then Eq.~(\ref{eq:FPF}) can be cast as 
\begin{equation}\label{eq:FPF2}
\begin{aligned}
\frac{\partial{\mathcal{F}}}{\partial{s}}-\sum_{i}{\bf K}'_{i}\lambda_{i}\frac{\partial\mathcal{F}}{\partial{\bf K}'_{i}}&=-\frac{{\bf K'}{\bf D}'{\bf K'}^{T}}{2}\mathcal{F},
\end{aligned}
\end{equation}
whose general solution is~\cite{Wang1945Brownian} 
\begin{equation}\label{eq:FPSGeneral}
\begin{aligned}
\mathcal{F}({\bf K}',s)&=f({\bf K}'{\bf \Lambda},0)e^{\left({\sum_{ij}\frac{1}{2}\frac{D_{ij}'}{\lambda_{i}+\lambda_{j}}{\bf K}_{i}'{\bf K}_{j}'}\right)},
\end{aligned}	
\end{equation} 
where ${\bf \Lambda}\equiv\text{diag}\left\{e^{\lambda_{1}s},\cdots,e^{\lambda_{6}s}\right\}$, with $\text{diag}\left\{\cdots\right\}$ means diagonal matrix.
The exact form of $f({\bf K}',0)$ is determined by the initial condition. 

Of special importance is the fundamental solution of Eq.~(\ref{eq:FPP}) or (\ref{eq:FPF}), i.e., the solution with an initial point charge distribution in phase space $\psi({\bf X},0)=\delta({\bf X}-{\bf X}_{0})$. The evolution of a general charge distribution can then be obtained based on this fundamental solution through superposition. With $\psi({\bf X},0)=\delta({\bf X}-{\bf X}_{0})$, we have $\mathcal{F}({\bf K},0)=e^{-i{\bf K}{\bf X}_{0}}$ and thus $\mathcal{F}({\bf K}',0)=e^{-i{\bf K}'{\bf P}{\bf X}_{0}}$, from which we have
\begin{equation}
f({\bf K}',0)=e^{-i{\bf K'PX}_{0}-\sum_{ij}\frac{1}{2}\frac{{\bf D}'_{ij}}{\lambda_{i}+\lambda_{j}}{\bf K}'_{i}{\bf K}'_{j}}.
\end{equation}
Denote ${\bf G}_{ij}\equiv-\frac{{\bf D}'_{ij}}{\lambda_{i}+\lambda_{j}}\left\{1-\text{exp}\left[(\lambda_{i}+\lambda_{j})s\right]\right\}$, from Eq.~(\ref{eq:FPSGeneral}) we then have 
\begin{equation}
\begin{aligned}
\mathcal{F}({\bf K}',s)&=e^{-i{\bf K}'{\bf \Lambda}{\bf PX}_{0}-\frac{1}{2}{\bf K'}{\bf G}{\bf K'}^{T}}.
\end{aligned}
\end{equation}
Denote ${\bf \sigma}\equiv{\bf P}^{-1}{\bf G}\left({\bf P}^{-1}\right)^{T}$, we then have 
\begin{equation}
\begin{aligned}
\mathcal{F}({\bf K},s)&=e^{-i{\bf K\hat{R}X}_{0}-\frac{1}{2}{\bf K}{\bf \sigma}{\bf K}^{T}}.
\end{aligned}
\end{equation}
Note that here ${\bf \hat{R}}(s,0)=e^{\int_{0}^{s}{\bf C}ds'}={\bf P}^{-1}{\bf \Lambda}{\bf P}$ is the deterministic transfer matrix with damping or antidamping. 
The above solution of $\mathcal{F}({\bf K},s)$ is the Fourier transform of a 6D Gaussian distribution given by 
\begin{equation}
\begin{aligned}
g({\bf X},{\bf X}_{0},s)&=\frac{\text{exp}\left[-\frac{1}{2}\left({\bf X}-{\bf \hat{R}}{\bf X}_{0}\right)^{T}{\bf \sigma}^{-1}\left({\bf X}-{\bf \hat{R}}{\bf X}_{0}\right)\right]}{(2\pi)^{3}\sqrt{\text{Det}({\bf \sigma})}}.
\end{aligned}
\end{equation}
The above result is the fundamental solution of Eq.~(\ref{eq:FPP}).
It means the average phase space coordinate and second moments of an initial point distribution evolve according to
\begin{equation}
\begin{aligned}
\langle{\bf X}\rangle(s)&={\bf \hat{R}}(s,0){\bf X}_{0},\\
\Sigma(s)&=\langle({\bf X}-{\bf \bar{X}})({\bf X}-{\bf \bar{X}})^{T}\rangle(s)={\bf \sigma}(s),
\end{aligned}
\end{equation} 
with
\begin{equation}
\begin{aligned}
{\bf \sigma}(s)
=\int_{0}^{s}{\bf \hat{R}}(s,s'){\bf D}(s'){\bf \hat{R}^{T}}(s,s')ds'.
\end{aligned}
\end{equation}
Note that 
\begin{equation}\label{eq:lmit1}
\begin{aligned}
\lim_{s\rightarrow0^{+}}g({\bf X},{\bf X}_{0},s)&=\delta({\bf X}-{\bf X}_{0}).
\end{aligned}
\end{equation}
If $\text{Re}(\lambda_{i})<0$ for $i=1\sim6$, then $\lim_{s\rightarrow+\infty}{\bf \hat{R}}(s,0)={\bf 0}$ and
\begin{equation}\label{eq:lmit2}
\begin{aligned}
\lim_{s\rightarrow+\infty}g({\bf X},{\bf X}_{0},s)&=\frac{\text{exp}\left(-\frac{1}{2}{\bf X}^{T}{\bf \sigma_{\infty}^{-1}}{\bf X}\right)}{(2\pi)^{3}\sqrt{\text{Det}({\bf \sigma_{\infty}})}},
\end{aligned}
\end{equation}
where $\sigma_{\infty}\equiv{\bf P}^{-1}{\bf G}_{\infty}\left({\bf P}^{-1}\right)^{T}$ with ${\bf G}_{ij,\infty}\equiv-\frac{{\bf D}'_{ij}}{\lambda_{i}+\lambda_{j}}$.

The evolution of a general beam distribution is then given by the superposition principle  
\begin{equation}\label{eq:superposition}
\psi({\bf X},s)=\int \psi({\bf X}_{0},0)g({\bf X},{\bf X}_{0},s)d{\bf X}_{0},
\end{equation} 
from which we have
\begin{equation}\label{eq:SFFDiffusion}
\mathcal{F}({\bf K},s)=\mathcal{F}({\bf K}{\bf \hat{R}},0)e^{-\frac{{\bf K\sigma K^{T}}}{2}}.
\end{equation}



With the above analysis done, the beam distribution can then be transported from element to element in a general coupled lattice. 
Generally we have
\begin{equation}\label{eq:SFFDiffusion7}
\mathcal{F}({\bf K},s)=\mathcal{F}({\bf K}{\bf \hat{R}},0)e^{-\frac{{\bf K\left(\int_{0}^{s}{\bf \hat{R}}(s,s'){\bf D}(s'){\bf \hat{R}}^{T}(s,s')ds'\right) K^{T}}}{2}},
\end{equation}
and
\begin{scriptsize}
	\begin{equation}\label{eq:SFFDiffusion72}
	\psi({\bf X},s)=\int\psi({\bf X}_{0},0)\frac{\text{exp}\left[-\frac{1}{2}\left({\bf X}-{\bf \hat{R}}{\bf X}_{0}\right)^{T}{\bf \sigma}^{-1}(s)\left({\bf X}-{\bf \hat{R}}{\bf X}_{0}\right)\right]}{(2\pi)^{3}\sqrt{\text{Det}({\bf \sigma}(s))}}d{\bf X}_{0}.
	\end{equation}
\end{scriptsize}
Equations~(\ref{eq:SFFDiffusion7}) and (\ref{eq:SFFDiffusion72}) are the main results of this section. Note that the above relations are accurate and hold for a general coupled linear lattice, arbitrary beam distribution in phase space, and does not require the beam to be Gaussian. 

\section{Electron Storage Ring Physics}

Now we apply the above analysis in electron storage ring physics. We will study the evolution of second moment matrix from $s_{1}$ to $s_{2}$ under the impact of damping and diffusion
{}
\begin{scriptsize}
	\begin{equation}\label{eq:BEnvelope}
	\begin{aligned}
	&{\bf \Sigma}(s_{2})=\int {\bf X}{\bf X}^{T}\psi({\bf X},s_{2})d{\bf X}\\
	&={\bf \hat{R}}(s_{2},s_{1}){\bf \Sigma}(s_{1}) {\bf \hat{R}^{T}(s_{2},s_{1})}+\int_{s_{1}}^{s_{2}}{\bf \hat{R}}(s_{2},s'){\bf D}(s'){\bf \hat{R}^{T}}(s_{2},s')ds'.
	\end{aligned}
	\end{equation}
\end{scriptsize}
This result is usually known as the beam envelope method~\cite{Ohmi1994Envelope}. We point out that Eqs.~(\ref{eq:SFFDiffusion7}) and (\ref{eq:SFFDiffusion72}) are more general and contain more information than the above evolution of second moments. In other words, we can derive Eq.~(\ref{eq:BEnvelope}) from Eq.~(\ref{eq:SFFDiffusion7}) or (\ref{eq:SFFDiffusion72}), but not the other way around.

In an electron storage ring, the equilibrium state repeats turn-by-turn, which means
\begin{scriptsize}
	\begin{equation}
	{\bf \Sigma}(s)={\bf \hat{M}}{\bf \Sigma}(s) {\bf \hat{M}^{T}}+\int_{s}^{s+C_{0}}{\bf \hat{R}}(s+C_{0},s'){\bf D}(s'){\bf \hat{R}}^{T}(s+C_{0},s')ds',
	\end{equation}
\end{scriptsize}
where ${\bf \Sigma}$ now is the equilibrium second moments matrix,  ${\bf \hat{M}}$ is the one-turn map evaluated at $s$ with damping around the ring considered. 
From this matrix equation, the equilibrium beam distribution can be solved.

Assuming that the damped one-turn map can be diagonalized as
$
{\bf \hat{M}}={\bf V}{\bf U}{\bf V}^{-1},
$
with ${\bf U}_{ii}=$ the eigenvalues of ${\bf \hat{M}}$, which means
$
{\bf \hat{M}}{\bf V}={\bf V}{\bf U}.
$
Denote
\begin{equation}
{\bf \mathcal{D}}(s)\equiv\int_{s}^{s+C_{0}}{\bf \hat{R}}(s+C_{0},s'){\bf D}(s'){\bf \hat{R}}^{T}(s+C_{0},s')ds',
\end{equation} 
then 
{}
\begin{equation}
\left({\bf V}^{-1}{\bf \Sigma}({\bf V}^{-1})^{T}\right)_{ij}(s)=\frac{\left[{\bf V}^{-1}{\bf \mathcal{D}}(s)({\bf V}^{-1})^{T}\right]_{ij}}{1-{\bf U}_{ii}{\bf U}_{jj}}.
\end{equation}
After getting ${\bf \Sigma}'\equiv{\bf V}^{-1}{\bf \Sigma}({\bf V}^{-1})^{T}$, we only need another transformation ${\bf V}{\bf \Sigma}'{\bf V}^{T}$ to get the equilibrium ${\bf \Sigma}$. In a linear lattice, the equilibrium beam distribution under the influence of damping and diffusion will tend to be Gaussian. For a Gaussian beam with $\langle{\bf X}\rangle=0$, we have $\mathcal{F}({\bf K},s)=e^{-\frac{{\bf K}{\bf \Sigma}(s){\bf K}^{T}}{2}}$ with ${\bf \Sigma}(s)=\langle {\bf X}(s){\bf X}^{T}(s)\rangle$ the second moments matrix of the beam. In this case, once we get the ${\bf \Sigma}$ matrix, the problem is solved. Under small damping approximation, the above result is consistent with Chao's SLIM formalism~\cite{Chao1979}.

\section{Structured Particle Beam}

Now we apply the SFF for the description of more structured particle beams, for example the laser-induced microbunching. We assume the initial beam distribution is Gaussian:
$
\psi_{0}({\bf X})=\frac{1}{(2\pi)^{3}\sqrt{\text{Det}({\bf \Sigma_{0}})}}\text{exp}\left(-\frac{1}{2}{\bf X}^{T}{\bf \Sigma_{0}^{-1}}{\bf X}\right)
$.

\subsection{One-Stage Energy Modulation}

Laser-induced energy modulation:
$
	\delta=\delta+A\sin(k_{L}z),
$
with $k_{L}=\frac{2\pi}{\lambda_{L}}$ the laser wavenumber.
Beam evolution after modulation:
$
		{\bf X}={\bf R}{\bf X},
$
where ${\bf R}$ is the linear transfer matrix of the magnet lattice. Denote:
$
		{\bf U}_{p}\equiv\left(
		\begin{matrix}
			0&
			0&
			0&
			0&
			pk_{L}&
			0
		\end{matrix}
		\right),\
		{\bf M}_{p}\equiv{\bf K}{\bf R}-{\bf U}_{p},\
		{\bf R}_{i6}\equiv\left(
		\begin{matrix}
			R_{16}&
			R_{26}&
			R_{36}&
			R_{46}&
			R_{56}&
			R_{66}
		\end{matrix}
		\right)^{T}.
$
Then the SFF after laser modulation and lattice transport is
\begin{equation}\label{eq:BFGeneral}
	\begin{aligned}
		\mathcal{F}({\bf K})		
		&=\sum_{p=-\infty}^{\infty}J_{p}\left(-{\bf K}{\bf R}_{i6}A\right)\text{exp}\left(-\frac{{\bf M}_{p}{\bf \Sigma}_{0}{\bf M}_{p}^{T}}{2}\right),
	\end{aligned}
\end{equation}
where $J_{p}$ is the $p$-th order Bessel function of the first kind. Note that damping and diffusion can also be taken into account following our analysis presented in last section.

Using the generalized Twiss matrices given in Ref.~\cite{OSCSSMB2024}, we can write ${\bf \Sigma}_{0}=\sum_{\kappa=I,II,III}\epsilon_{\kappa}{\bf T}_{\kappa}$,
where ${\bf \epsilon}_{\kappa}$ and ${\bf T}_{\kappa}$ are the eigen emittance the generalized Twiss matrix of the $\kappa$-th eigenmode at the beginning. For a periodic system like a storage ring, if ${\bf R}={\bf T}_{0}^{m}$ with ${\bf T}_{0}$ the symplectic one-turn map, which means the beam traverses the ring for $m$ revolutions after the laser modulation, then
\begin{scriptsize}
	\begin{equation}
	\begin{aligned}
	{\bf M}_{p}{\bf \Sigma}_{0}{\bf M}_{p}^{T}
	&=\sum_{\kappa=I,II,III}\epsilon_{\kappa}\left({\bf K}{\bf T}_{\kappa}{\bf K}^{T}-2{\bf K}{\bf R}{\bf T}_{\kappa}{\bf U}_{p}^{T}+{\bf U}_{p}{\bf T}_{\kappa}{\bf U}_{p}^{T}\right).
	\end{aligned}
	\end{equation}
\end{scriptsize}
The above analysis can be applied in the proof-of-principle experiment of steady-state microbunching (SSMB)~\cite{Ratner2010SSMB,Deng2021Experimental,Kruschinski2024Confirming}, for example to investigate the microbunching wavefront tilt and off-axis coherent radiation generation after laser modulation arising from transverse-longitudinal coupling~\cite{DengSpringer2024}.


\subsection{Double-Stage Energy Modulations}
First-stage laser modulation:
$
	\delta=\delta+A_{1}\sin(k_{1}z),
$
with $k_{1}$ the wavenumber of first laser.
Beam evolution after first modulation:
$
		{\bf X}={\bf R}_{1}{\bf X}.
$
Second-stage laser modulation:
$
	\delta=\delta+A_{2}\sin(k_{2}z+\phi).
$
Beam evolution after second modulation:
$
		{\bf X}={\bf R}_{2}{\bf X}.
$
Denote:
$
 		{\bf U}_{2}\equiv\left(
 		\begin{matrix}
 			0&
 			0&
 			0&
 			0&
 			m_{2}k_{2}&
 			0
 		\end{matrix}
 		\right),\
 		{\bf M}_{2}\equiv{\bf K}{\bf R}_{2}-{\bf U}_{2},\
 		{\bf U}_{1}\equiv\left(
 		\begin{matrix}
 			0&
 			0&
 			0&
 			0&
 			m_{1}k_{1}&
 			0
 		\end{matrix}
 		\right),\
 		{\bf M}_{1}\equiv{\bf M}_{2}{\bf R}_{1}-{\bf U}_{1},\
 		{\bf R}_{1i6}\equiv\left(
 		\begin{matrix}
 			R_{1,16}&
 			R_{1,26}&
 			R_{1,36}&
 			R_{1,46}&
 			R_{1,56}&
 			R_{1,66}
 		\end{matrix}
 		\right)^{T},\
 		{\bf R}_{2i6}\equiv\left(
 		\begin{matrix}
 			R_{2,16}&
 			R_{2,26}&
 			R_{2,36}&
 			R_{2,46}&
 			R_{2,56}&
 			R_{2,66}
 		\end{matrix}
 		\right)^{T}.
$
 Following steps similar to above, the final SFF is
 \begin{equation}\label{eq:BFEEHG}
 	\begin{aligned}
 		\mathcal{F}({\bf K})
 		&=\sum_{p_{1}=-\infty}^{\infty}\sum_{p_{2}=-\infty}^{\infty}J_{p_{1}}\left(-{\bf M_{2}}{\bf R}_{1i6}A_{1}\right)e^{ip_{2}\phi}\\
 		&J_{p_{2}}\left(-{\bf K}{\bf R}_{2i6}A_{2}\right)\text{exp}\left\{-\frac{{\bf M}_{1}{\bf \Sigma}_{0}{\bf M}_{1}^{T}}{2}\right\}.
 	\end{aligned}
 \end{equation}
 For a periodic system like a storage ring, if ${\bf R}_{1}={\bf T}_{0}^{m_{1}}$, ${\bf R}_{2}={\bf T}_{0}^{m_{2}}$  with ${\bf T}_{0}$ the one-turn map, then 
 \begin{scriptsize}
 	\begin{equation}
 	\begin{aligned}
 	&{\bf M}_{1}{\bf \Sigma}_{0}{\bf M}_{1}^{T}=\sum_{\kappa=I,II,III}\epsilon_{\kappa}\left({\bf K}{\bf T}_{\kappa}{\bf K}^{T}+{\bf U}_{2}{\bf T}_{\kappa}{\bf U}_{2}^{T}+{\bf U}_{1}{\bf T}_{\kappa}{\bf U}_{1}^{T}\right.\\
 	&\left.\ \ \ \ -2{\bf K}{\bf R_{2}}{\bf T}_{\kappa}{\bf U}_{2}^{T}-2{\bf K}{\bf R_{2}R_{1}}{\bf T}_{\kappa}{\bf U}_{1}^{T}+2{\bf U_{2}R_{1}}{\bf T}_{\kappa}{\bf U}_{1}^{T}\right).
 	\end{aligned}
 	\end{equation}
 \end{scriptsize}
 Similar procedures can be applied to derive the SFF for even more multiple-stage modulations.


Representative examples of the above single-stage and double-stage energy modulation-based laser-induced microbunching schemes are HGHG~\cite{Yu1991} and EEHG~\cite{Stupakov2009}, respectively. Based on the above derivations and more calculations, we have proposed to do EEHG at the MLS storage ring using the SSMB proof-of-principle experiment setup~\cite{Deng2021Experimental,Kruschinski2024Confirming}. We can use different revolution numbers to play the role of different $R_{56}$ in the two-stage dispersions of EEHG. For example, we can fire the first-shot laser and wait for $2n+1$ revolutions, and then we fire the second-shot laser and then wait for 2 revolutions. Finally we can generate high harmonic bunching at the $m\times n$-th laser harmonic.





%

\section{Summary}\label{sec:summary}
%
%
%
%

In this paper, spectral form function (SFF) has been presented as a tool to study beam dynamics.  More interesting work on this is ongoing and will be reported in the future. The author thanks Alex Chao for helpful discussion. 





\end{document}